\documentclass[a4paper,12pt]{article}

\usepackage[height=8.85in,width=6.45in]{geometry}
\usepackage{fancybox}
\usepackage{amsmath,amssymb}
\usepackage[T1]{fontenc}
\usepackage{comment}
\usepackage{cite}
\usepackage{hyperref}
\usepackage{xcolor}

\def\CC{{\cal C}}

\def\CH{{\cal H}}

\def\CT{{\cal T}}

\usepackage{slashed}

\def\pin{\mathop{\rm pin}}

\def\SL{\mathop{\rm SL}}
\def\tr{\mathop{\rm tr}}

\def\CC{{\cal C}}

\def\CH{{\cal H}}

\def\CT{{\cal T}}

\newcommand{\bC}{\mathbb{C}}
\newcommand{\bR}{\mathbb{R}}
\newcommand{\bZ}{\mathbb{Z}}

\def\SL{\mathrm{SL}}
\def\tr{\mathop{\mathrm{tr}}\nolimits}

\def\sT{\mathsf{T}}
\def\sC{\mathsf{C}}
\def\sP{\mathsf{P}}
\def\sR{\mathsf{R}}

\def\sCPT{\mathsf{CPT}}
\def\sCRT{\mathsf{CRT}}

\def\beq#1\eeq{\begin{align}#1\end{align}}

\usepackage{framed}
\definecolor{shadecolor}{rgb}{0.90,0.90,0.90}

\usepackage{times}
\usepackage{courier}
\usepackage{mathtools}
\numberwithin{equation}{section}

\let\bar\overline

\newcommand{\cT}{\mathcal{T}}

\def\CC{\mathbb{CC}}
\def\MO{\mathbb{MO}}

\newcommand{\bra}[1]{ \langle {#1} | }
\newcommand{\ket}[1]{ | {#1} \rangle }
\newcommand{\et}[1]{  {#1} \rangle }

\begin{document} 

\begin{titlepage}

\begin{flushright}
IPMU-16-0158
\end{flushright}

\vfill

\begin{center}

{\Large\bfseries More on time-reversal anomaly of 2+1d topological phases}

\vskip 1cm
Yuji Tachikawa and Kazuya Yonekura
\vskip 1cm

\begin{tabular}{ll}
  & Kavli Institute for the Physics and Mathematics of the Universe, \\
& University of Tokyo,  Kashiwa, Chiba 277-8583, Japan
\end{tabular}

\vskip 1.5cm

\textbf{Abstract}

\end{center}

\vskip1cm

\noindent

We prove an explicit formula conjectured recently by Wang and Levin for the anomaly of time-reversal symmetry in 2+1 dimensional fermionic topological quantum field theories.
The crucial step is to determine the crosscap state in terms of the modular S matrix and $\sT^2$ eigenvalues, generalizing the recent analysis by Barkeshli et al. in the bosonic case.
\vfill

\end{titlepage}
\tableofcontents

\section{Introduction and heuristic derivation}
The 3+1 dimensional fermionic topological superconductors with $\sT^2=(-1)^F$
are classified by $\bZ_{16}$, as shown by many arguments such as dynamics
\cite{Fidkowski:2013jua,Wang:2014lca,KitaevCollapse,Metlitski:2014xqa,Morimoto:2015lua,Seiberg:2016rsg,Tachikawa:2016xvs,Witten:2016cio},
cobordism group~\cite{Kapustin:2014dxa} and the anomaly of the 2+1 dimensional systems which appear on the boundary of the topological superconductors~\cite{Witten:2015aba,Hsieh:2015xaa}.

In a previous work~\cite{Tachikawa:2016cha}, the present authors have discussed a way to compute the anomaly of the edge theory when it is a
topological quantum field theory (TQFT). The procedure is that we consider the state $\ket{\CC}$ on a torus $T^2$ which is created on the boundary of the
3-dimensional manifold [crosscap $\times$ circle], and then compute the eigenvalue of the modular transformation $T \in \SL(2,\bZ)$ of the torus acting on $\ket{\CC}$,
\beq
T\ket{\CC}=\exp(\frac{2\pi i \nu}{16}) \ket{\CC},\label{eq:centralformula}
\eeq
where $\nu \in \bZ_{16}$ is the anomaly of the time reversal symmetry.

The crosscap state $\ket{\CC}$ must satisfy several consistency conditions discussed in \cite{Tachikawa:2016cha},
and the different crosscap states satisfying the consistency conditions correspond to different extensions to nonorientable manifolds of a given theory on orientable manifolds.
In that way, the known values of the anomaly $\nu \in \bZ_{16}$ were exactly reproduced.
However, it was not clear in \cite{Tachikawa:2016cha} how to use the information of the eigenvalue of the square $\sT^2$ of the time-reversal symmetry $\sT$
acting on quasi-particles, which is quite important
to distinguish different theories~\cite{Metlitski:2014xqa,Seiberg:2016rsg,Witten:2016cio}.

In an independent work by Wang and Levin~\cite{Wang:2016qkb}, a very explicit formula for the anomaly 
$\exp(\frac{2\pi i \nu}{16})$ was conjectured, which is given in our notation as follows:
\beq
\exp(\frac{2\pi i \nu}{16}) = \frac{1}{D} \sum_{p} \eta_p d_p e^{-2 \pi i h_p} . \label{eq:WL}
\eeq 
Here, the sum runs over quasiparticles $p$ such that $\sT p =p$ or $\sT p=pf$,
$D$ is the total quantum dimension of the system,
$d_p$, $h_p$, $\eta_p$ are
 the quantum dimension,
 the topological spin,
 and the eigenvalue of $\sT^2$ on the quasiparticle on $p$ multiplied by $-i$ if $\sT p=pf$.
 Here the $f$ is a neutral fermion which can escape to the bulk of the topological superconductors.
 The $\sT p=pf$ is possible because the quasi-particles are anyons for which the notion of boson/fermion is rather ambiguous. 
 
The origin of that formula was not discussed in that paper; rather, the conjecture was supported by many examples. 
The formula is very important because it immediately tells us which theory can appear on the surface of a given topological superconductor.

In this paper we derive the formula above by combining the result of \cite{Tachikawa:2016cha} together with an explicit form of the crosscap state 
\begin{equation}
\ket{\CC} = S \sum_{p} \eta_p  \ket{p},\label{eq:CCx}
\end{equation} where $S$ is the modular S matrix.
The expression above for the bosonic case was done in the case of bosonic systems by 
Berkeshli, Bonderson, Cheng, Jian and Walker \cite{Barkeshli:2016mew} and announced in 
 a talk by Barkeshli \cite{KITPTalk}, and we derive \eqref{eq:CCx} by extending  their results to the fermionic case.

\paragraph{Heuristic derivation}
Because our derivation of the above formula is very technical, 
we would like to give a less technical (but at the same time less rigorous) explanation here which hopefully makes the physical meanings of the various notions used in this paper clearer. 
The $2+1$ dimensional theories considered in this paper appears on the edge of $3+1$ dimensional topological superconductors as mentioned above.
We can detect the quantity $\nu \in \bZ_{16}$ characterizing the topological superconductor as follows~\cite{Shiozaki:2016zjg}.\footnote{
The authors would like to thank K.~Shiozaki for helpful discussions.}
Let us separate a space into two regions A and B where the region $A$ is a 3 dimensional ball whose boundary is a sphere $\partial A =S^2$.
Then consider the Hilbert spaces $\CH_A$ and $\CH_B$ on them.
What is nontrivial about the ground state $\ket{\Omega}$ of the topological superconductors (and more generally symmetry protected topological phases) is that
if we write $\ket{\Omega} = \sum_i\ket{\Omega_A}_i \otimes \ket{\Omega_B}_i$ for $ \ket{\Omega_{A,B}}_i \in \CH_{A,B}$, the symmetry actions on 
each of $ \ket{\Omega_{A}}_i$ and $ \ket{\Omega_{B}}_i$ do not give genuine representations; rather, they have nontrivial phase factors and are projective representations~\cite{Chen:2011pg}. 
Because of this, if we try to trace out $\CH_B$, there appears nontrivial edge theory on the boundary between two regions
which reproduce the nontrivial phase under the action of the symmetry, which is $\exp( 2\pi i \nu/16)$ for the topological superconductors.

In the present case, the relevant symmetry is the time reversal $\sT$ which is related to the inversion $\mathsf{Inv}: \vec{x} \mapsto  - \vec{x}$ by $\sCPT$ theorem.
Then, the nontrivial phase under the action of the symmetry $\mathsf{Inv}$ appears as~\cite{Shiozaki:2016zjg}
\beq
\exp(\frac{2\pi i \nu}{16}) \sim \sum_{\ket{\alpha} \in \CH_{ S^2}} \bra{\alpha} \mathsf{Inv} \ket{\alpha} \label{eq:quasisum}
\eeq
where the sum is over the states in the Hilbert space $\CH_{ S^2}$ of the edge theory on the boundary $S^2=\partial A$.
The $\mathsf{Inv}$ can be written as 
$
\mathsf{Inv} = \sR_{S^2} \cdot \exp( - i \pi  \mathsf{J}_{S^2})
$
where $\sR_{S^2} $ is a reflection in the $x$-direction, and $\mathsf{J}_{S^2}$ is the angular momentum around the $x$-axis.

The edge theory Hilbert space $\CH_{ S^2}$ contains the ground state and also several excitations of quasiparticles.
In particular, we can consider states $\ket{p,p'}$ which contain a quasiparticle $p$ at $\vec{x}=(1,0,0)$ and $p'$ at $\vec{x}=(-1,0,0)$.
For this state to exist, the $p'$ needs to be the anti-particle $\bar{p}$ of $p$ up to a neutral fermion $f$ which can escape to the bulk of the topological superconductors.
The angular momentum of $p$ and $p'$ add up to give $ \mathsf{J}_{S^2} = h_p - h_{p'}$ where $h_p$ is the spin of $p$.
The $\sR_{S^2} $ acts as $\sR_{S^2}  \ket{p,p'} \sim  \ket{ \sR(p'),\sR(p)} $ where $\sR(p)$ is the quasiparticle related to $p$ by the symmetry $\sR$. 
This means that only the states with $p'=\sR(p)$ contribute to the sum, 
and $h_p - h_{\sR(p)}=2h_p$ because the reflection $\sR$ changes the sign of the spin of a particle.
The action of $\sR$ on $\sR(p)$, namely $\sR^2(p)$, gives a phase factor which we denote as $\eta_p$. By $\sCPT$ theorem,
the $\eta_p$ is related to $\sT^2_p$. Furthermore, the absolute value of the contribution of $ \ket{p,\sR(p)}$ to the sum \eqref{eq:quasisum} should be proportional to 
the ``dimension" of subspace spanned by them in the Hilbert space, i.e., the quantum dimension $d_p$. Notice that the $d_p$ appears instead of $(d_p)^2$ because
the two quasiparticles $p$ and $p'$ are exchanged by $\sR_{S^2}$ and hence there is only a single world-line of $p$ in the spacetime 
representation of $\bra{p,p'}  \mathsf{Inv} \ket{p,p'}$.
Therefore,  \eqref{eq:quasisum} becomes
\beq
\exp(\frac{2\pi i \nu}{16}) \sim \sum_{p} e^{-2 \pi i  h_p} \eta_p d_p
\eeq
which is the formula \eqref{eq:WL} up to a positive overall factor $D$.


The rest of the paper gives a more rigorous (but not completely mathematically rigorous) explanation of the formula. It consists of two sections. 
In Sec.~\ref{sec:2} we derive the explicit formula \eqref{eq:CCx}
and show the relation  between $\eta_p$  and $\sT^2_p$.
Then  in Sec.~\ref{sec:3}  we deduce the Wang-Levin formula \eqref{eq:WL} 
from \eqref{eq:centralformula} and \eqref{eq:CCx}  by a simple manipulation.

\section{The crosscap state and the time-reversal squared}\label{sec:2}
Let us consider the manifold $\MO  _A\times S^1_B$, where $\MO  _A$ is the M\"obius strip, connecting the boundary $S^1_A$ and the crosscap given by
\beq
\MO  _A=\{ (x,\theta) \in [-1, 1] \times \bR ; (x,\theta) \sim (-x,\theta+\pi) \}  . \label{eq:mobius}
\eeq 
It has the crosscap at $x=0$ and the boundary $S_A^1 = \partial \MO_A$ at $x=1$.
The $S^1_B$ is a copy of the circle. This manifold has the boundary torus $T^2=\partial (\MO  _A\times S^1_B)=S_A^1 \times S_B^1$.\footnote{Note that $T^2$ is the torus while $\sT^2$ is the time-reversal squared.}
The spin structure on $S_A^1$ is automatically R (i.e., periodic) as discussed in \cite{Tachikawa:2016cha}, whereas the spin structure on $S_B^1$ can be
taken arbitrarily. We take it to be NS (i.e., anti-periodic) for later convenience. 

There is the state $\ket{\CC}$ on the boundary torus which is created by the manifold $\MO  _A\times S^1_B$.
We call it the crosscap state.
We want to determine this crosscap state $\ket{\CC}$ by using the information of $\sT^2$.
For this purpose, we try to expand the state in a complete basis of the Hilbert space on $T^2$.

Two complete bases can be constructed as follows. Let us consider $S^1_A \times D_B(p)$, where $D_B(p)$ is a two-dimensional disk $D_B$ 
with the line operator $p$ inserted at the center of the disk times $S^1_A$. 
The boundary of this manifold is again $T^2=S_A^1 \times S_B^1$.
We can also consider a manifold $D_A(p) \times S^1_B$ in which the role of the $A$-cycles and $B$-cycles of $T^2$ are exchanged from the $S^1_A \times D_B(p)$.
These two are related by the modular transformation $S \in \SL(2,Z)$ acting on the torus.
Let $\ket{p}$ be the state on $T^2$ created by $D_A(p) \times S^1_B$. Then the state created by $S^1_A \times D_B(p)$
is given by $S\ket{p}$ where $S$ now acts on the Hilbert space.
More precisely,
when the spin structure is taken into account, the $S$ may change the spin structure, and hence can be 
considered as a map from one Hilbert space to another. Namely, if $\CH_{s_1,s_2}$ ($s_{1,2}=$ NS or R) is the Hilbert space on 
$T^2=S_A^1 \times S_B^1$ with spin
structure $s_1$ on the A-cycle $S_A^1$ and $s_2$ on the B-cycle $S_B^1$, then $S$ is a map $S: \CH_{s_1,s_2} \to  \CH_{s_2,s_1} $.

The $\ket{\CC}$ is in $\CH_{\text{R},\text{NS}}$. If we take the spin structures of $D_A(p) \times S^1_B$ as $(\text{NS,R})$,
then the states $\ket{p}$ are in $\CH_{\text{NS},\text{R}}$ and hence $S\ket{p} \in \CH_{\text{R},\text{NS}}$. 
Therefore, we can expand $\ket{\CC}$ as
\beq
\ket{\CC} = \sum_{p} \eta_p S \ket{p},\label{eq:CC}
\eeq
where $\eta_p$ are coefficients.
These coefficients $\eta_p$ can be computed as 
\beq
\eta_p= \bra{p}S^{-1}\ket{\CC}. \label{eq:coefficients}
\eeq

Geometrically, this quantity can be interpreted as follows. We mentioned above that the $S$-transformation of $D_A(p) \times S^1_B$ is given by
$S^1_A \times D_B(p)$. Therefore, \eqref{eq:coefficients} can be interpreted as the partition function on the manifold
which is obtained by gluing $\MO  _A\times S^1_B$ and 
$S^1_A \times D_B(p)$ along the boundary $T^2=S_A^1 \times S_B^1$
with the line $p$ inserted. We denote this manifold as $X(p)$.

For the moment, we forget about the line operator $p$.
To see how the glued manifold looks like, let us take the oriented double cover of $\MO_A$ in \eqref{eq:mobius} as
\beq
\tilde{\MO}_A = [-1, 1] \times {S}^1_A 
\eeq 
Then the oriented double cover $\tilde{X}$ of $X$ is given by gluing the $S^1_A \times D_B$ and also another manifold obtained by spatial reflection of $S^1_A \times D_B$
to $\tilde{\MO}_A \times S_B = {S}^1_A \times  [-1, 1]  \times S_B^1$. The result of the gluing is given by
\beq
\tilde{X}= S^1_A \times S^2_B,
\eeq
where $S^2_B$ a 2-dimensional sphere obtained by gluing $ D_B$ and its reflection to the boundaries of $[-1, 1]  \times S_B^1$.

We describe $S^1_A $ and $ S^2_B$ as
\beq
S^1_A=\{\theta; \theta \sim \theta+2\pi\},~~~S^2_B=\{ \vec{n}=(n_x,n_y,n_z); |\vec{n}|=1 \}.
\eeq
Now we define a diffeomorphism $\sigma$ which acts on $S^1_A \times S^2_B$ as
\beq
\sigma: (\theta, n_x,n_y,n_z) \mapsto (\theta+\pi, -n_x,n_y,n_z). 
\eeq
Then we see that the manifold $X$ is given by dividing $\tilde{X}$ by the equivalence relation
defined by $\sigma$. We simply write it as
\beq
X =[S^1_A \times S^2_B]/\sigma.
\eeq

If we restore the line $p$, then we have two lines wrapping around $S_A^1$ on $\tilde{X}$ because we took two copies of $S^1_A \times D_B(p)$, one of which was  reflected.
The fact that one of them is acted by the spatial reflection means that the two line operators are given by $p$ and $\sR p$, where $\sR p$
is the type of line operator which is related to $p$ by the reflection symmetry.
In summary, we get
\beq
X(p)=[S^1_A \times S^2_B(p, \sR p)]/\sigma
\eeq
where $S^2_B(p, \sR p)$ is the $S^2_B$ with the line $p$ and $\sR p$ inserted at the north pole $(n_x,n_y,n_z)=(1,0,0)$ and
the south pole $(-1,0,0)$, respectively.

Let $\CH(p, p')$ be the Hilbert space on $S^2$ with two time-like lines $p$ and $p'$ inserted at the north and south pole, respectively.
Also let $\sR_{S^2}$ be the operator $ \sR_{S^2}: \CH(p, p') \to \CH(\sR p', \sR p) $
which implements the diffeomorphism $(n_x,n_y,n_z) \mapsto (-n_x,n_y,n_z)$. The definition of $\sR_{S^2}$ has an ambiguity by $(-1)^F$
which corresponds to the two $\pin^+$ structures on $X$. This goes back to the $\pin^+$ structures on $\MO  _A \times S_B^1$.
Restricting to the case $p'=\sR p$, the $\sR_{S^2}$ maps $\CH(p, \sR p) $ to itself.
Then the $\eta_p= \bra{p}S^{-1}\ket{\CC}$ is finally given by
\beq
\eta_p=\tr_{\CH(p, \sR p)}( \sR_{S^2}).\label{eq:etap}
\eeq
This trace gives the partition function on $X(p)$.
We interpret this $\eta_p$ as the quantity $\tilde{\CT}^2_p$ appearing in \cite{Wang:2016qkb},
which is the eigenvalue of $\sT^2$ acting on the quasi-particle $p$, multiplied by $-i $  when $\sT p=pf$.

Let us first see
some consequences of the definition \eqref{eq:etap}:
\begin{enumerate}
\item The Hilbert space $\CH(p, p')$ is zero unless $p'$ is the CRT conjugate of $p$ up to transparent fermion $f$, i.e.,
$p' =\bar{p}$ or $f \bar{p}$. 
Therefore, $\eta_p$ is zero unless $\sT p =  p$ or $f  p$, where $\sT p := \sR \bar{p}$ is the line operator related to $p$ by the time reversal symmetry.
In other words, $p$ must be self-conjugate under the time reversal up to $f$.
\item When $\CH(p, \sR p)$ is nonzero, it is always one-dimensional. Also, $\sR_{S^2}$ satisfies $(\sR_{S^2})^2=1$ because we are considering the 
$\pin^+$ theories. 
Therefore, $\eta_p =\pm 1$.
\item If we change the $\pin^+$ structure to the opposite one, we replace $\sR_{S^2} \to \sR_{S^2}'=\sR_{S^2} (-1)^F$
and correspondingly we get $\eta_p \to \eta'_p$, where $\eta'_p=\eta_p$ for $ \sT p =  p$ and $\eta'_p= - \eta_p$ for $\sT p = f p$.
\item For a trivial operator $1$ we get $\eta_1=1$. For two operators $p$ and $\bar{p}$ which are CRT conjugate of each other, we get $\eta_p=\eta_{\bar{p}}$
as a consequence of the CRT theorem.
\end{enumerate}
In the last claim $\eta_p=\eta_{\bar{p}}$, the relevant CRT transformation is given as follows. First, we define $\sC$ to be trivial.
Second, we consider $\tilde{\sR}_{S^2} : (n_x,n_y,n_z) \mapsto (n_x,-n_y,n_z)$ which is different from $\sR_{S^2}$ defined above.
This new $\tilde{\sR}_{S^2}$ fixes the positions of two points $( \pm 1,0,0)$. 
\def\sCtildeRT{\mathsf{C}\tilde{\mathsf{R}}\mathsf{T}}
Then we define $\sCtildeRT :=\tilde{\sR}_{S^2} \sT$. 
This $\sCtildeRT$ is a map $\sCtildeRT: \CH_{p, p'} \to \CH_{\bar{p},\bar{p}'}$, and
also satisfies $(\sCtildeRT)^\dagger \sR_{S^2} (\sCtildeRT)=\sR_{S^2}$. Thus the result $\eta_p=\eta_{\bar{p}}$ follows.

Now let us relate $\eta_p$ and $\sT^2_p$. 
Unfortunately we still do not have a full understanding of how to interpret the time-reversal $\cT$ acting on a single quasiparticle purely from the point of view of the topological quantum field theory, 
but we will try our best.
Let $\ket{p, q} \in \CH(p, q)$ be the unique state on $S^2_B(p,q)$.
The basic idea is that there are two quasi-particles $p$ and $q=\sT \bar{p}$ and hence $\sT^2_p \in \bC$ appears as
\beq
\sT  \ket{p, \sT \bar{p}}  \sim \sT^2_p \ket{\sT p, \bar{p}}.
\eeq
However, for this purpose, we need a way to compare $\ket{\sT p, \bar{p}}$ and $ \ket{p, \sT \bar{p}} $ by exchanging the two particles.

For this purpose, we have two operators which can be used as isomorphisms between different Hilbert spaces.
One isomorphism is $\sCtildeRT$ defined above. 
The other is $i^{F^2}  e^{\pi  iJ} $, where $e^{\pi  iJ} :  (n_x,n_y,n_z) \mapsto  (-n_x,-n_y,n_z)$
is the $\pi$ rotation around the $n_z$ axis. The reason that we put $i^{F^2} $ is because we get the relation $(i^{F^2}  e^{\pi  iJ} )^2=1$ so that the braiding phase
of exchanging the two quasi-particles by $e^{\pi  iJ}$ is canceled by $i^{F^2}$.
We also have $[\sCtildeRT, i^{F^2}  e^{\pi  iJ}]=0$ and $ (\sCtildeRT )^2=1$ for CRT transformations\footnote{The usual CPT transformation which is familiar 
in four space-time dimensions has $(\sCPT)^2=(-1)^F$. However, CRT has $(\sCRT)^2=1$, because $\sR$ flips only one coordinate while $\sP$ flips three coordinates.}.
Then, there are natural isomorphisms between the Hilbert spaces as
\beq
\sCtildeRT ~:~& \CH(p,q)  \to \CH(\bar{p},\bar{q}) ,\\
i^{F^2}  e^{\pi  iJ} ~:~& \CH(p,q) \to \CH(q,p).
\eeq 
The former is antilinear while the latter is linear.

Using these isomorphisms, we define the $\sT^2$ eigenvalue of the quasiparticle $p$ in the language of the topological quantum field theory by the formula
\beq
 \sT^2_p \ket{p, \sT \bar{p}}  :=
i^{F^2}  e^{\pi  iJ} \cdot \sCtildeRT \cdot  \sT  \ket{p, \sT \bar{p}}  \label{A}
\eeq
Indeed, under these transformations, a single quasi-particle transforms as
\beq
&p_{(n_x=1)} \xrightarrow{~\sT~} \sT p_{(n_x=1)}   \xrightarrow{\sCtildeRT} \sT \bar{p}_{(n_x=1)}   \xrightarrow{i^{F^2}  e^{\pi  iJ} } \sT \bar{p}_{(n_x=-1)} \nonumber \\
&\xrightarrow{~\sT~}  \bar{p}_{(n_x=-1)}   \xrightarrow{\sCtildeRT}  p_{(n_x=-1)}   \xrightarrow{i^{F^2}  e^{\pi  iJ} } p_{(n_x=1)}  \label{eq:experience}
\eeq
Therefore, the single quasi-particle experiences $\sT$ twice, up to the isomorphism $ i^{F^2}  e^{\pi  iJ} \cdot \sCtildeRT$.
Notice that the $\sT$ is preserved under this isomorphism, 
\beq
  \sT (i^{F^2}  e^{\pi  iJ} \cdot \sCtildeRT)=  (i^{F^2}  e^{\pi  iJ} \cdot \sCtildeRT) \sT,\label{eq:preserveT}
\eeq
because $ \sCtildeRT \cdot \sT = (-1)^F \sT \cdot  \sCtildeRT$, $e^{\pi  iJ} \sT = \sT e^{\pi  iJ}$  and $ i^{F^2} \sT = \sT(-i)^{F^2}  = (-1)^F  \sT i^{F^2}$.
If we had not included the factor $i^{F^2}$, there would have been an extra factor $(-1)^F$ on the right-hand-side of \eqref{eq:preserveT} and
hence the interpretation of \eqref{A} as giving $\sT_p^2$ would have been impossible.

The $\sCtildeRT$ has been defined above as $\tilde{\sR}_{S^2} \sT$. Furthermore, we have the relation $e^{\pi  iJ} = \tilde{\sR}_{S^2} \sR_{S^2}$.
Therefore, we get
\beq
i^{F^2}  e^{\pi  iJ} \cdot \sCtildeRT \cdot  \sT =i^{F^2} \sR_{S^2} .\label{B}
\eeq
The statistics of the state on $\CH_{p, \bar{p}}$ is bosonic, as can be seen by pair annihilating $p$ and $\bar{p}$.
In the same way, the statistics of the state on $\CH_{p, f\bar{p}}$ is fermionic because a single $f$ remains after pair annihilating $p$ and $\bar{p}$. Therefore, 
\beq
 \bra{p, \sT \bar{p}} i^{F^2} \ket{p, \sT \bar{p}} = \left\{ \begin{array}{ll} 
 1 & \text{if }\sT p =p \\ 
 i & \text{if }\sT p =fp  
 \end{array} \right.\label{C}
\eeq
Combining \eqref{A}, \eqref{B} and \eqref{C}, we get the desired result $\sT^2_p=\eta_p$ for $\sT p=p$ and $\sT^2_p = i \eta_p$ for $\sT p=f p$.

Before moving on, we remark the following. In \cite{Tachikawa:2016cha}, it was noted in Sec.~4.4 there that when two crosscap states are related by $\ket{\CC}_{qX}=B(q)\ket{\CC}_X$ where $B(q)$ is the quasiparticle $q$ wrapped around $S^1_B$, the values of $\sT^2_p$ in the theory $qX$ and in the theory $X$ differ by the braidin phase of $p$ and $q$. This naturally follows from the identification of $\eta_p$ and $\sT^2_p$, since acting by $B(q)$ changes $\eta_p$ by the braiding phase of $p$ and $q$.

\section{The formula for the time reversal anomaly}\label{sec:3}
Now we derive the formula for the time reversal anomaly proposed in \cite{Wang:2016qkb}.
In  a previous paper by the present authors \cite{Tachikawa:2016cha}, the anomaly $\nu \in \bZ_{16}$ is given by the eigenvalue of the modular transformation $T \in \SL(2,\bZ)$
acting on the crosscap state $\ket{\CC}$,
\beq
T\ket{\CC} = \exp(\frac{2\pi i \nu}{16}) \ket{\CC}.
\eeq
We note that
$
\bra{1} S^{-1} \ket{\CC} = \sum_{p} \eta_p \bra{1}  \et{p}=1,
$
 Therefore, we get 
$
\exp(\frac{2\pi i \nu}{16})=\bra{1} S^{-1}T \ket{\CC}.
$

Let us compute the right-hand side. Recall that we have the relations $T\ket{p}_{\text{NS},\text{R}}=e^{2 \pi i h_p} \ket{p}_{\text{NS},\text{NS}}$,
where $\ket{p}_{s_1, s_2}$ is the state with spin structure $s_1$ on the A-cycle and $s_2$ on the B-cycle, and $h_p$ is the spin of $p$.
The $s_1$ must be correlated with the type of $p$, that is, $s_1=$ NS or R depending on whether $p$ is NS or R line operator in the terminology of \cite{Tachikawa:2016cha}.
The $T\ket{p}_{\text{NS},\text{R}}=e^{2 \pi i h_p} \ket{p}_{\text{NS},\text{NS}}$ needs a clarification which will be discussed below.
To be explicit, we attach a subscript $(s_1,s_2)$ to states to make it clear which spin structures the state is defined on.

Now, note that $\ket{\CC}=S\sum_p \eta_p\ket{p}=S^{-1} \sum_p\eta_p \ket{p}$ because $S^2\ket{p}=\ket{\bar{p}}$ and $\eta_{\bar{p}}=\eta_p$. Then 
\begin{equation}
\begin{aligned}
{}_{\text{NS},\text{R}} \bra{1} S^{-1}T \ket{\CC}_{\text{NS},\text{R}}  
&=   {}_{\text{NS},\text{R}} \bra{1} S^{-1} T S^{-1} \sum_{p} \eta_p \ket{p}_{\text{NS},\text{R}} \\
&=   {}_{\text{NS},\text{R}} \bra{1} T^{-1} S T^{-1} \sum_{p} \eta_p \ket{p}_{\text{NS},\text{R}} 
=\sum_{p} \eta_p \cdot (S_{\text{NS},\text{NS}})_{0,p} e^{-2 \pi i h_p} ,
\end{aligned}
\end{equation}
where we used $(S^{-1}T)^3=1$ in the second equality.

We are almost done. The quantity $(S_{\text{NS},\text{NS}})_{0,p}$ is the matrix element of $S$ in the $(\text{NS},\text{NS})$ sector.
This matrix element is given by $d_p/ D$, where $d_p=(S_{\text{NS},\text{NS}})_{0,p}/(S_{\text{NS},\text{NS}})_{0,0}$ is the quantum dimension
of the line operator $p$ (i.e., the expectation value of the line operator on a trivial knot  $S^1$), 
and $D=1/(S_{\text{NS},\text{NS}})_{0,0}=\sqrt{\sum_a (d_a)^2}$ is the total dimension (i.e., the inverse of the $S^3$ partition function).

In summary, we get the formula
\beq
\exp(\frac{2\pi i \nu}{16}) = \frac{1}{D} \sum_{p} \eta_p d_p e^{-2 \pi i h_p} . \label{eq:theformula}
\eeq
This is exactly the formula proposed in \cite{Wang:2016qkb}, up to the remark we make in the next paragraph.
There, $\eta_p$ was denoted as $\tilde{\CT}^2_p $ and $-2\pi h_p$ was denoted as $\theta_p$.

The formula \eqref{eq:theformula} looks different from the eq.(10) of \cite{Wang:2016qkb} by a factor of $\sqrt{2}$.
The reason is as follows.
As an example, let us consider the semion-fermion theory. This theory is usually supposed to contain four operators (and corresponding states),
$1$, $f$, $s$ and $s'=sf$, where $f$ is the transparent fermion, $s$ is the semion and $s'=sf$ is the anti-semion.
However, after specifying the spin structure, there are actually only two states as a spin TQFT on the torus $T^2$.
Schematically, when $s_2=\text{NS}$ (and $s_1=$NS automatically\footnote{The spin structure in the A-cycle direction $s_1$ must be $\text{NS}$ for $1$, $f$, $s$ and $sf$.
See \cite{Tachikawa:2016cha} for the case of $s_1=$R.}),
the two states are given by
\beq
\ket{1}_{s_2=\text{NS}  } \sim \ket{1} + \ket{f},~~~~~\ket{s}_{s_2=\text{NS}}  \sim \ket{s} + \ket{f s}.
\eeq
The two states of the case $s_2=\text{R}$ are given by
\beq
\ket{1}_{s_2=\text{R}  } \sim \ket{1} - \ket{f},~~~~~\ket{s}_{s_2=\text{R}}  \sim \ket{s} - \ket{f s}.
\eeq
The sum \eqref{eq:theformula} is given only over these two states $\ket{1}_{s_2=\rm NS}$ and $\ket{s}_{s_2=\rm NS}$.
With this understanding, the formula \eqref{eq:theformula} in this semion-fermion theory is evaluated as
\beq
\exp(\frac{2\pi i \nu}{16}) = \frac{1}{\sqrt{2}}  (1-\eta_s i)
\eeq
where we have used $h_s=1/4$. The $\eta_s$ is given by $\eta_s= \pm 1$ depending on two different semion-fermion theories ($\text{SF}_+$ and $\text{SF}_-$),
and we get $\nu= \pm 2$. This reproduces the known result.

In general, $h_p$ and $h_{fp}$ differ by $1/2$, but we mentioned that once we fix the spin structure, $p$ and $fp$ are the same, i.e., $p_\text{NS}=(fp)_\text{NS}$.
So the definition of $h_p$ is ambiguous.
However, the combination $\eta_{p}e^{2 \pi i h_p}$ is unambiguous if $\eta_{fp}= - \eta_{p}$.
For example, the $\sT^2=(-1)^F$ eigenvalue on $f$ is $\eta_f=-1$, which is consistent with the above claim.

The formula \eqref{eq:theformula} was already checked in many examples \cite{Wang:2016qkb}, so we do not repeat it here.

\section*{Acknowledgements}
The authors thank Chenjie Wang for informing them the paper \cite{Wang:2016qkb} and the talk \cite{KITPTalk}
which are crucial in the present work. They also thank K.~Shiozaki for helpful discussions.
The work of Y.T. is partially supported in part by JSPS Grant-in-Aid for Scientific Research No. 25870159.
The work of K.Y. and Y.T. is supported by World Premier International Research Center Initiative
(WPI Initiative), MEXT, Japan.

\addtocontents{toc}{\protect\setcounter{tocdepth}{1}}

\bibliographystyle{ytphys}
\baselineskip=.9\baselineskip
\let\bbb\bibitem\def\bibitem{\itemsep1pt\bbb}
\bibliography{ref}

\end{document}